\documentclass[a4paper]{jpconf}
\usepackage{graphicx}
\usepackage{amsmath,amssymb}
\begin{document}
\title{A family of ternary decagonal tilings}

\author{Nobuhisa Fujita}

\address{Institute of Multidisciplinary Research for Advanced Materials, Tohoku University, Sendai 980-8577, Japan}

\ead{nobuhisa@tagen.tohoku.ac.jp}

\begin{abstract}
A new family of decagonal quasiperiodic tilings are constructed by the
use of generalized point substitution processes, which is a new substitution
formalism developed by the author
[N. Fujita, {\it Acta Cryst. A 65, 342 (2009)}].
These tilings are composed of three prototiles: an acute rhombus, a regular
pentagon and a barrel shaped hexagon.
In the perpendicular space, these tilings have windows with fractal
boundaries, and the windows are analytically derived as the fixed sets
of the conjugate maps associated with the relevant substitution rules.
It is shown that the family contains an infinite number of local
isomorphism classes which can be grouped into several symmetry classes
(e.g., $C_{10}$, $D_5$, etc.).
The member tilings are transformed into one another through collective
simpleton flips, which are associated with the reorganization in the
window boundaries.
\end{abstract}


\section{Introduction}


The standard projection formalism \cite{debruijn81,mackay82,kramer84,duneau85}
for constructing quasiperiodic point sets had evolved into the so called
{\it dualization method}, which can be used to construct extensive
{\it quasiperiodic tilings}; a complete account of the dualization method
can be found in \cite{schlottmann93} and references cited therein.
The generated tilings however are restricted to those whose windows are given
as convex polytopes. Recall that some quasiperiodic tilings are known to have
windows which have fractal boundaries, take for instance the dodecagonal
square-triangle tilings \cite{baake92,smith93}. Such tilings, including many
unknowns too, are unlikely to be constructed by way of projection because of
the following difficulties. (i) Handling directly the broad range of fractal
windows, which also fulfil the non-trivial constraints for tilings with a
moderately small number of prototiles, is generally rather difficult.
(ii) Even if the window is known, it is practically difficult to avoid minute
numerical errors associated with the fine structures of the window when
projecting the higher-dimensional structure down to the physical space.

These difficulties however are not relevant if instead a substitution
(or an inflation) formalism is used for construction. Recall that various
substitution rules for constructing
quasiperiodic tilings with fractal windows can be found in the literature
\cite{stampfli86,zobetz92,luck93,godreche93,hermisson97}.
We describe a substitution rule for a tiling by the following two steps:
(S1) expansive similarity transformation of the tiling and (S2) division of
the expanded tiles into tiles of the original size. For a deterministic
tilings, it is commonly assumed that the division of every expanded tile is
determined uniquely by the local surroundings of that tile. Moreover,
division rules for neighbouring expanded tiles must not compete with each
other because tiles are not allowed to overlap.
To search for a substitution rule of a new tiling, trial and error has been
commonplace. This uneconomic situation has led me to propose a more
systematic way for composing substitution rules for generating quasiperiodic
tilings. The new method is called {\it generalized point substitution
processes} (GPSPs) \cite{fujita09}, in which Step S2 is divided into
two sub-steps: (S2a) decoration of the expanded tiles with points which
would be the candidate positions of the vertices of the next tiling and
(S2b) choose an appropriate subset of the candidate positions so that the
chosen points can be connected by unit edges to generate the next tiling.
These sub-steps operate in a local manner so that they conform to the local
uniqueness of the division Step S2.

In the subsequent two sections, the basic information about the decagonal
Bravais module and the GPSP scheme is summarized. Then in Sec.4 the scheme
is extensively applied to construct the new family of ternary decagonal
quasiperiodic tilings. Note that Similar tilings have already been
discussed in \cite{fujita09}. 
The aim is to somewhat extend the method used there so that it can handle
a broader class of tilings.
It turns out that the present method can generate tilings whose point
groups are $D_5$, $C_5$, $C_2$, $D_1$ and $C_1$, in addition to $C_{10}$
which has been the only point group for the ternary tilings discussed in
\cite{fujita09}. In the final section, we introduce a randomization
procedure into the GPSP scheme, where we obtain further variety of tilings
with partial disorder. The effects of simpleton flips in the tilings are
also considered.

\section{Decagonal Bravais module}

The five crystallographic axes for decagonal systems can be represented
by the five unit vectors, $\boldsymbol{e}_j=(\cos{(j\theta)},\sin{(j\theta)})$
with $j=0, 1, ..., 4$ and $\theta=2\pi/5$. The first four members are taken as
the basis set for indexing the lattice points in the system.
The basis set generates a Z-module of rank four called the decagonal Bravais
module \cite{rokhsar87}, denoted by the symbol $\boldsymbol{\Lambda}_{10}$.
$\boldsymbol{\Lambda}_{10}$ has the point symmetry of the dihedral group
$D_{10}$, while it has an important property of scaling invariance
$\tau\boldsymbol{\Lambda}_{10}=\boldsymbol{\Lambda}_{10}$, where
$\tau=(1+\sqrt{5})/2$ is the golden mean.

The two-dimensional axis vectors in $\Bbb{E}^{\parallel}$ are lifted into
the four-dimensional Euclidean hyperspace $\Bbb{E}_4$ as
$\boldsymbol{\epsilon}_j=(\boldsymbol{e}_j, \boldsymbol{e}^{\perp}_{j})$,
where the complementary components are defined by
$\boldsymbol{e}^{\perp}_j:=\boldsymbol{e}_{2j\;({\rm mod}\;5)}$ where $j=$
0, 1, ..., 4.
The first four members of $\{\boldsymbol{\epsilon}_j\}$ are the generators
of the decagonal lattice $\boldsymbol{L}_{10}$. The orthogonal projection of
$\boldsymbol{L}_{10}$ onto $\Bbb{E}^\parallel$ gives nothing but the
decagonal Bravais module $\boldsymbol{\Lambda}_{10}$ \cite{niizeki89}.
The two-dimensional
perpendicular space $\Bbb{E}^\perp$ is defined as the orthogonal complement
to $\Bbb{E}^\parallel$ in $\Bbb{E}_4$; that is,
$\Bbb{E}_4=\Bbb{E}^\parallel \oplus \Bbb{E}^\perp$.
Then $\boldsymbol{L}_{10}$ can be projected onto $\Bbb{E}^{\perp}$ as well,
generating the conjugate module $\boldsymbol{\Lambda}_{10}^{\perp}$.
Since the orthogonal projections from $\boldsymbol{L}_{10}$ to both
$\boldsymbol{\Lambda}_{10}$ and $\boldsymbol{\Lambda}_{10}^\perp$ are
bijections, one can introduce a natural bijection $\hat{\pi}$ between the
modules;
$\boldsymbol{\Lambda}_{10}^{\perp}=\hat{\pi}\boldsymbol{\Lambda}_{10}$.

\section{Generalized point substitution processes}

We make a restricted  use of the term {\it planar tilings} for disjoint
coverings of the plane by copies of a finite number of polygonal prototiles
while fulfilling edge-to-edge condition.
From such a tiling, a point set can always be defined by taking the set
of vertices. In contrast, however, it is not always possible to define a
tiling from a point set. Let us confine ourselves to the case when points
from the set can be connected with uncrossed unit edges given by the
vectors $\boldsymbol{e}_j$ ($j=0, 1, ...,4$) and these edges define a
tessellation of the plane by a finite kinds of polygons. Then the point
set can be called {\it unit connective}. It is further assumed that the
vertex set $\boldsymbol{\Sigma}_\mathcal{T}$ of a tiling $\mathcal{T}$ is
a subset of the Bravais module $\boldsymbol{\Lambda}_{10}$.

The window $\boldsymbol{W}_\mathcal{T} (\subset \Bbb{E}^\perp)$ is so 
defined that $\boldsymbol{\Sigma}_\mathcal{T}$ is the orthogonal projection of
$(\Bbb{E}^\parallel+\boldsymbol{W}_\mathcal{T})\cap\boldsymbol{L}_{10}$ onto
$\Bbb{E}^\parallel$. \footnote{Here the $+$ symbol implies that
 $\boldsymbol{A}+\boldsymbol{B}\equiv
\{a+b | ^\forall{a}\in\boldsymbol{A},^\forall{b}\in\boldsymbol{B}\}$.}
Then the image of $\boldsymbol{\Sigma}_\mathcal{T}$ in $\Bbb{E}^\perp$,
denoted as $\boldsymbol{\Sigma}_\mathcal{T}^\perp:=
\hat{\pi}\boldsymbol{\Sigma}_\mathcal{T}$,
is a dense subset of the window $\boldsymbol{W}_\mathcal{T}$, which on
the other hand should be included in the closure of
$\boldsymbol{\Sigma}_\mathcal{T}^\perp$, i.e. 
$\boldsymbol{\Sigma}_\mathcal{T}^\perp\subset
\boldsymbol{W}_\mathcal{T}\subseteq
\overline{\boldsymbol{\Sigma}_\mathcal{T}^\perp}$.

A GPSP \cite{fujita09} is described as a three-step process for composing a
substitution rule:
(G1) {\it expansive similarity transformation of the original tiling by the
ratio $\sigma = \tau^n$ ($\tau$ : the Pisot unit, or the golden mean for the
decagonal case),}
(G2) {\it replication of the basic motif $\boldsymbol{S}$ on every vertex} and
(G3) {\it elimination of excess points, so that the remaining points
are unit connective.}
Steps G1 and G2 together constitute a {\it point inflation rule}, which
has been introduced for constructing a quasiperiodic point set
\cite{niizeki08}, whereas Step G3 is an crucial step for constructing
quasiperiodic tilings \cite{fujita09}.

\section{RPH tilings}

RPH tilings have three prototiles, R (a rhombus with acute angles of
36$^\circ$), P (a regular pentagon) and H (a barrel shaped hexagon).
Every edge corresponds to one of the five unit vectors
\{$\boldsymbol{e}_j$\} defined above. In \cite{fujita09}, two GPSPs
have been identified as substitution rules for RPH tilings. These GPSPs are
the mirror images of one another and are connected through simpleton
flips involving three adjacent tiles, R, P and H. The two GPSPs can be
applied in a mixed and arbitrary order so that an infinite number of
tilings with the point group $C_{10}$ can be obtained.
The conjugate maps for these GPSPs are defined in the perpendicular space,
and the windows with fractal boundaries for the relevant RPH tilings are
derived analytically as the fixed sets of these conjugate maps.

\begin{figure}[b]
\begin{center}
\begin{minipage}{17pc}
\begin{center}
\includegraphics[width=12pc]{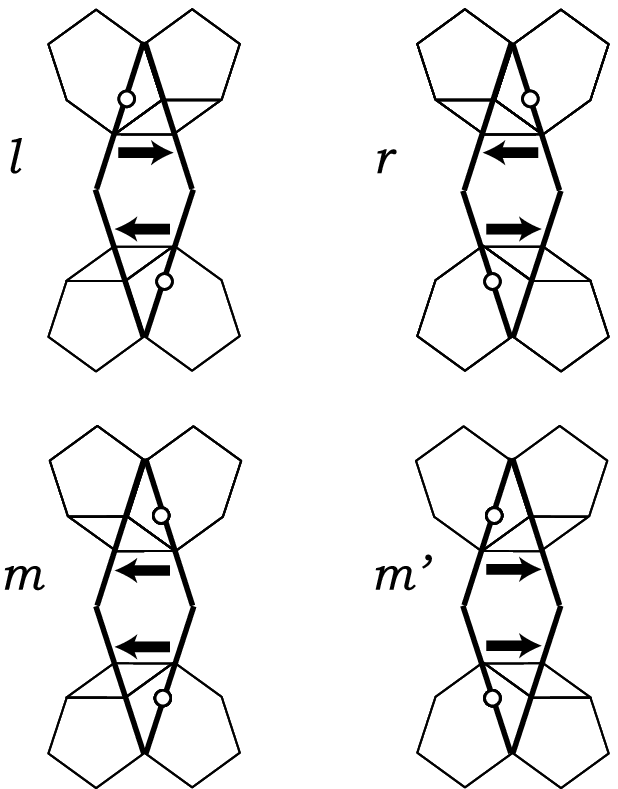}
\end{center}
\caption{\label{fig:elimination} Points can be eliminated in four distinct
ways from an expanded rhombus. The eliminated points are represented as
open circles. The choice of a side for each acute angle can be represented
by an arrow.}
\end{minipage}\hspace{2pc}
\begin{minipage}{17pc}
\begin{center}
\includegraphics[width=13.5pc]{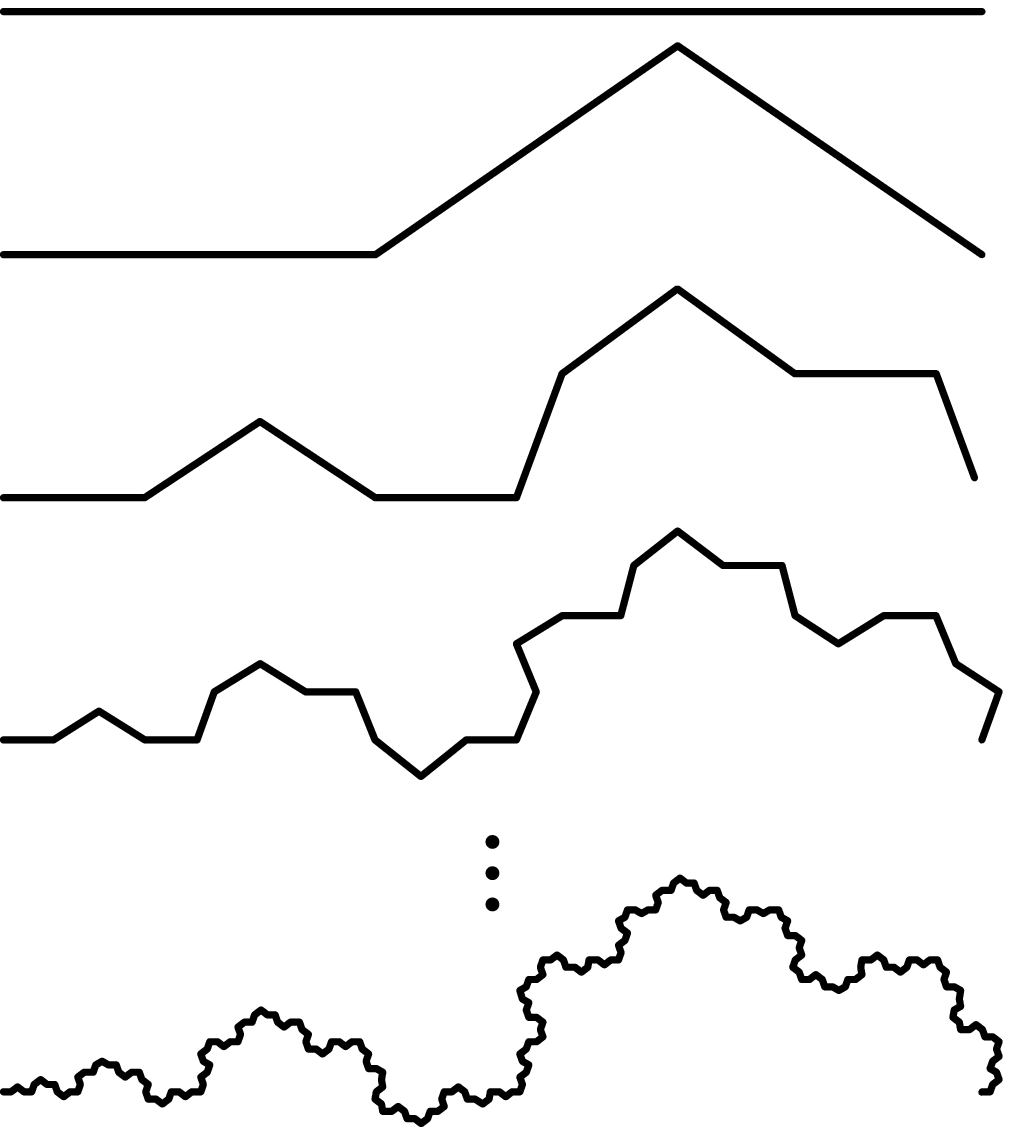}
\end{center}
\caption{\label{fig:fractaldim} A 1/20 sector of the boundary of the
window is constructed in a recursive manner. The corresponding window
(Fig.\ref{fig:symmetries}, upper left) is for the case when only the
{\it l}-type elimination rule is applied.
}
\end{minipage} 
\end{center}
\end{figure}

Let us provide some information about the two GPSPs for RPH tilings.
The scaling ratio for the expansion step G1 is taken as the square of
the golden mean $\tau$. The basic motif $\boldsymbol{S}$ for Step G2
is a centred decagon, with the representative indices being [0000]
(centre) and [1100] (the decagonal shell). The first two steps, G1 and G2,
would generate the candidate positions for the vertices of the next RPH
tiling. These positions are superfluous for the vertices of an RPH tiling,
so that an excess part should be eliminated at Step G3. To be specific,
either of the two points that lie on every acute angle of expanded rhombi
at the unit distance from the apex is to be eliminated. There are four
ways to eliminate such points from an expanded rhombus, as shown in
Fig.\ref{fig:elimination}; one can assign an arrow to each of the acute
angles according to the choice of a point that is eliminated.

\begin{figure}[b]
\begin{center}
\includegraphics[width=28pc]{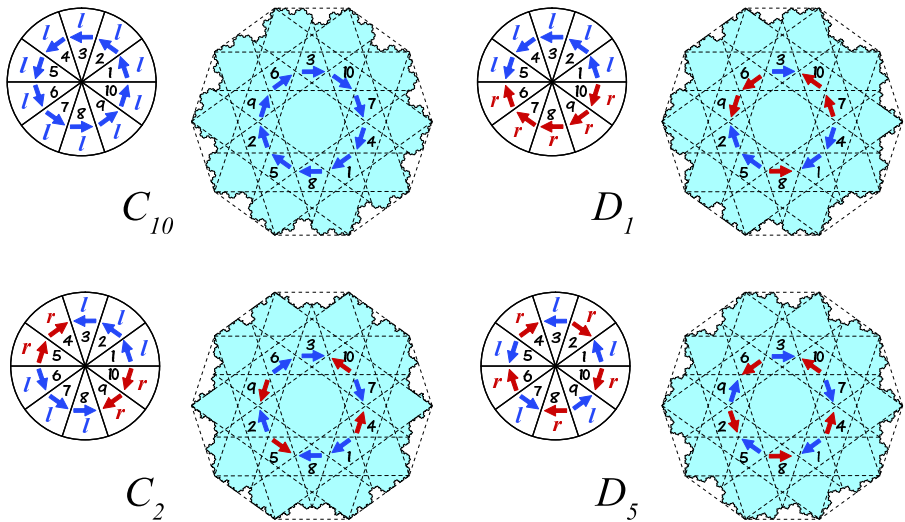}
\caption{Four different ways to assign arrows to acute angles depending
on their directions are represented by the wheel diagrams. Their point
group symmetries are as indicated. The windows for the RPH tilings
generated by successively applying the relevant GPSPs are depicted.}
\label{fig:symmetries}
\end{center}
\end{figure}

Since the two elimination rules labelled {\it l} and {\it r} (signifying
the left- and right-handed chiralities, respectively) in
Fig.\ref{fig:elimination} are invariant under the two-fold rotation of
the rhombus, either of these rules can be applied uniquely to all the
expanded rhombi in a single iteration. It has been argued that the
relevant two chiral GPSPs could be applied in an arbitrary order, resulting
in a family of RPH tilings with the point group $C_{10}$ \cite{fujita09}.
For the left-handed GPSP, the arrows assigned to all the acute angles are
symmetrically equivalent. Let us represent this situation as the wheel
diagram shown in the upper left part of Fig.\ref{fig:symmetries}.

It turns out that the window for an RPH tiling generated by the two
GPSPs has a fractal boundary with the point symmetry $C_{10}$.
The window can be analytically represented as the limit figure obtained
by successively applying the conjugate maps associated with the GPSPs.
A conjugate map operates in $\Bbb{E}^\perp$ and comprises the following
three steps:
(D1) {\it contractive similarity transformation by the ratio $\bar{\sigma}$,
which is the algebraic conjugate to $\sigma$ and
$|\bar{\sigma}|=1/|\sigma|<1$,}
(D2) {\it replication of the contracted figure onto every point
of the conjugate motif $\boldsymbol{S}^\perp=\hat{\pi}\boldsymbol{S}$} and
(D3) {\it subtraction of an excess part of the resulting figure
corresponding to the elimination step G3 in the physical space.}
Note that, for decagonal RPH tilings, the algebraic conjugate to
$\sigma=\tau^2$ is given by $\bar{\sigma}=1/\tau^2$ for Step D1.
It has been shown \cite{fujita09} that Step D3 for RPH tilings can be
described as a carving process of a small portion near the boundary of
the figure obtained after Step D2. Note that ten strips are indicated by
the broken lines in the window diagrams in Fig.\ref{fig:symmetries}.
Corresponding to the arrows for the ten acute angles in the wheel
diagram, arrows can be associated with the ten strips. Then the aft-end
of each strip (tail of the arrow) is so carved that it would coincide
with the fore-end of the strip through the $\tau$-translation.

In Fig.\ref{fig:fractaldim}, a 1/20 sector of the boundary of the window
is shown for first several iterations of the conjugate map for the left-handed
GPSP. At each iteration, each line segment is replaced with an array of
three segments, which are $1/\tau^2$ times shorter than the original one.
As the conjugate map is repeated infinitely, the boundary would be a
generalized von Koch curve whose fractal dimension is simply calculated
as $dim(\partial\boldsymbol{W}_\mathcal{T})=\ln3/\ln(\tau^2)\doteqdot 1.1415$.

It is also tempting to allow different elimination rules (i.e., arrow
designations) for each expanded rhombus. Let us confine ourselves to the
case when the choice of an elimination rule for each expanded rhombus
only depends on the orientation of the rhombus. The choice for all the
orientations can be specified by an wheel diagram. There are
$2^{10}(=1024)$ distinct arrow designations for wheel diagrams. In
Fig.\ref{fig:symmetries}, four wheel diagrams as well as the corresponding
windows are shown. Different shapes of the windows imply that the relevant
tilings belong to distinct local isomorphism classes. The global point
symmetry of a tiling (which is the same as that of the window) can be
inferred from the point symmetry of the relevant wheel diagram.

\begin{figure}[b]
\begin{center}
\includegraphics[width=28pc]{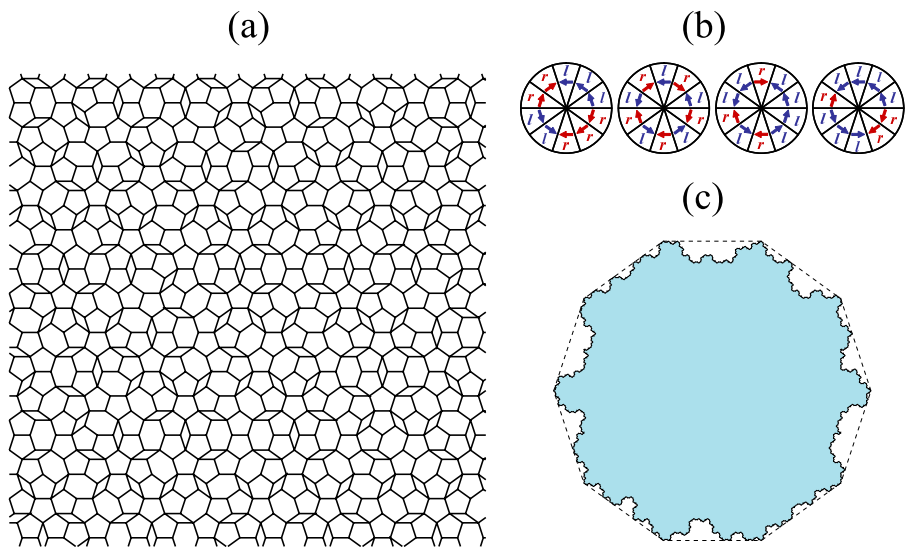}
\caption{\label{fig:sample1} Part (a) shows a square patch of the RPH tiling
generated by successively applying the four-cycle of GPSPs that are
represented in Part (b). Part (c) depicts the window for the tiling.
There is no point group symmetry except for the trivial one.}
\end{center}
\end{figure}

It is readily seen that a tiling with a mirror symmetry could be
generated only if the elimination rule $m$ or $m'$ (see
Fig.\ref{fig:elimination}) would takes place for at least one orientation
of expanded rhombi. Since these elimination rules do not allow a
two-fold axis, a two-fold rotational symmetry is incompatible with
a mirror symmetry. Therefore, a global mirror symmetry exists only when
a two-fold rotational symmetry is absent; the only allowed point groups
are $D_5$ and $D_1$. In addition, $C_2$ and $C_1$ may also be the point
group of a wheel diagram with arrows. Moreover, if different GPSPs are
applied in a mixed and arbitrary order, the resulting point group would
be a common subgroup to the point groups of the relevant GPSPs;
$C_5$ is allowed in this way. Obviously, most of the $1024$
possible wheel diagrams have the point group $C_1$.
In Fig.\ref{fig:sample1}, an RPH tiling generated by four-cycles of GPSPs
is demonstrated along with the corresponding window.

In \cite{fujita09}, statistics of the local tile arrangements in RPH
tilings have been analysed by two substitution matrices associated
with the GPSPs of the types $l$ and $r$. It has been shown that the
relative frequencies of the three prototiles are invariably given by
$n(R):n(P):n(H)=1:2:1/\tau$. An alternative way to prove the statistics
is to take the advantage of the four-dimensional description of the
structures.

Let us denote by $v$ the density of the vertices of an RPH tiling. Then
it is given by the formula: $v=w/\Omega_4$, where $w$ is the area of the
window and $\Omega_4$ the four-dimensional volume of the primitive unit
cell of the decagonal lattice $\boldsymbol{L}_{10}$.
For the family of RPH tilings, $w=2\sqrt{5} a$ and $\Omega_4=5\sqrt{5}/4$,
leading to $v=(8/5)a$, where $a=\sin(\pi/5)=\tau^{-1/2}5^{1/4}/2$ is the
area of an R tile. The point density is related to the densities of R, P
and H tiles in the following way:
\begin{eqnarray}
n(R) + \frac{3}{2} n(P) + 2 n(H) &=& \frac{8}{5}a.
\end{eqnarray}
Recall also that an RPH tiling is a projection of a `puckered' net which
is embedded in $\boldsymbol{L}_{10}$ along the physical space. The
puckered net can be projected onto a lattice plane spanned by two of the
five axis vectors $\boldsymbol{\epsilon}_i$ and $\boldsymbol{\epsilon}_j$
($i\ne j$). A specific shape (R, P or H) and orientation of tiles in the
physical space corresponds to a specific contribution to the area of the
projected image. The total area of the projected image is proportional to
that of the tiling in $\Bbb{E}^\parallel$. There are ten lattice planes
onto which the tiling can be mapped, and accordingly there are ten
equations for the densities of the relevant tiles. By appropriately
adding the left and right hand sides of these equations, one derives the
following two equations for the density of the three prototiles:
\begin{eqnarray}
n(R) + \frac{1}{2} n(P) + n(H) &=& \frac{4}{5} a,\\
\frac{3}{2} n(P) + 2 n(H) &=& \frac{4}{5} \tau a.
\end{eqnarray}
From Eqs.(1-3), it turns out that the densities for the three prototiles are
\begin{equation}
  n(R)=\frac{4}{5\tau^2}a,~
  n(P)=\frac{8}{5\tau^2}a,~
  n(H)=\frac{4}{5\tau^3}a.
\end{equation}
The right hand sides of Eqs.(1-3) may be modified if there exists a linear
phason strain, which however is assumed to be absent in the present report.

\section{Random RPH tilings}

The 1024 GPSPs as described above can be applied in an arbitrary order,
resulting in an unlimited number of RPH tilings or, more precisely, local
isomorphism classes; these tilings still maintain perfect quasiperiodicity.
Let us now lift the constraint on the uniqueness of the elimination rule
for all the expanded rhombi in the same orientation. If a new GPSP,
in which the elimination rule for each expanded rhombus is chosen at random,
is used at each iteration, a member of a stochastic ensemble of RPH tilings
will be obtained. It will still fulfil the edge-to-edge condition.
The number of possible patterns within a patch of diameter $R$ will
diverge exponentially as $R$ is increased. Such a randomized GPSP can be
thought of as a realization of the concept of {\it random substitution rules}
proposed by Gummelt \cite{gummelt08}, who gave the example of randomized
Penrose/T\"ubingen-triangle tilings.

In every GPSP described so far, irrespective of it being randomized or not,
the elimination step G3 takes a subset of a larger point set, which is
generated through Steps G1 and G2. It follows that by repeating the latter
two steps only, a unique super-set can be defined from which the vertices
of any RPH tiling originate. The super-set is the set of vertices of the
{\it para-Penrose tiling} \cite{niizeki08}, whose window is a moth-eaten
version of the regular decagon; see Fig.4-no.1 in \cite{niizeki08}.
Then for any of the randomized RPH tilings described above, points close
to the boundary of this window is partially removed. The partial
occupancies in certain triangular regions close to the boundary of the
window are the same as the deterministic case as given in the end of the
caption of Fig.8 in \cite{fujita09}; the only difference lies in that
in the present case points are removed in a stochastic manner and that
there is no clear boundary between occupied and unoccupied parts.
In all cases the main part of the window is fully occupied except the
partially occupied boundary regions, hence the quasiperiodicity is almost
maintained. The relatively small amount of randomness will cause weak
diffuse components superposed in the diffraction spectra.

The elimination rules for expanded rhombus as shown in Fig.1 can be
mutually connected through simpleton flips involving three tiles, R, P and
H. Every flip is caused by a hop of a vertex by a distance of $1/\tau$ in
the parallel direction to the relevant arrow. This corresponds to a
transport of a point in the perpendicular space by a distance of $\tau$
along the associated strip; see Fig.3. Moreover, all the RPH tilings we
have introduced so far are mutually connected through collective simpleton
flips of the same kind, because such flips can be repeated indefinitely so
that the boundary regions of the window can be reorganized.

We have also performed a preliminary Monte Carlo study in which simpleton
flips are excited stochastically when there is the combination of three tiles
R, P and H. It has turned out that the resulting tilings are genuinely
random, where the quasiperiodic order in the initial tiling has been
completely dissolved while only the orientational order has been maintained.

Finally, it may be helpful to redirect the reader to some physical
counterparts. The same kind of simpleton flips as discussed above has been
observed in situ in a {\it d}-Al-Cu-Co quasicrystal at a temperature of 1123 K
\cite{edagawa00}.
Furthermore, the centres of atomic clusters in a high-temperature phase
of {\it d}-Al-Ni-Co quasicrystal can be described rather nicely by the
vertices of an RPH tiling \cite{hiraga91,niizeki94}.
Therefore, the family of RPH tilings may serve as templates for physical
models of decagonal quasicrystals.



\end{document}